\documentclass[journal]{IEEEtran}
\usepackage{graphicx}
\begin{document}
\title{First results from the FPGA/NIOS Adaptive FIR Filter Using Linear Prediction
Implemented in the AERA Radio Stations
to Reduce Narrow Band RFI for Radio Detection of Cosmic Rays}
%
%

\author{Zbigniew~Szadkowski,~\IEEEmembership{Member,~IEEE,}, D.~G{\l}as, 
C.~Timmermans, T.~Wijnen,\\ for the Pierre Auger Collaboration
\thanks{Manuscript received June 03, 2014.} 
\thanks{Zbigniew~Szadkowski is with the University of \L{}\'od\'{z}, Department of Physics and Applied Informatics, 
Faculty of High-Energy Astrophysics, 90-236 \L{}\'od\'{z}, Poland,
(e-mail: zszadkow @kfd2.phys.uni.lodz.pl, phone: +48 42 635 56 59).}%
\thanks{Dariusz~G{\l}as is the PhD student with the University of \L{}\'od\'{z}, Department of Physics and Applied Informatics, 
Faculty of High-Energy Astrophysics, 90-236 \L{}\'od\'{z}, Poland,}%
\thanks{Charles~Timmermans, Thei~Wijnen are with Nikhef, Science Park, Amsterdam, Netherlands,  and IMAPP, Radboud University Nijmegen, Netherlands}
}
\maketitle
\thispagestyle{empty}

\begin{abstract}
The FPGA/NIOS$\textsuperscript{\textregistered}$ FIR filter based on linear
prediction (LP) to suppress radio frequency interference (RFI) has been installed
in several radio stations in the Auger Engineering Radio Array (AERA) experiment. 
AERA observes coherent radio emission from extensive air showers induced 
by ultra-high-energy cosmic rays to make a detailed study of the development
of the electromagnetic part of air showers. Radio signals provide  complementary 
information to that obtained from Auger surface detectors, which are predominantly sensitive to
the particle content of an air shower at the surface. The radio signals
from air showers are caused by the coherent emission due to geomagnetic
and charge-excess processes. These emissions can be observed in the
frequency band between 10 - 100 MHz. However, this frequency range
is significantly contaminated by narrow-band RFI and other human-made
distortions. A FIR filter implemented in the FPGA logic segment of
the front-end electronics of a radio sensor significantly improves
the signal-to-noise ratio.
Theoretical calculations show a high efficiency of this filter for mono-carrier
as well as for standard FM radio contaminations.
The laboratory tests, performed on the Altera$\textsuperscript{\textregistered}$ 
Cyclone$\textsuperscript{\textregistered}$ V DK-DEV-5CEA7N
development kit confirmed the theoretical expectations.

In this paper we present first results of the efficiency of the adaptive LP FIR filter, 
deployed in real AERA station on pampas, with a comparison to the currently used
IIR notch filter with constant coefficients.
The coefficients for the linear predictor are dynamically refreshed and calculated 
in a Voipac PXA270M  ARM processor, which is implemented on a daughter-board placed  
in the same digital unit as the FPGA.  The laboratory tests confirms the stability of the filter. Using constant
LP coefficients the suppression efficiency remains the same for hours, 
which corresponds to more than $\bf 10^{12}$ clock cycles.
We compared in real conditions several variants of the LP FIR filter with various
lengths and various coefficients widths (due to fixed-point representations in the FPGA logic)
with the aim to minimize the power consumption for the radio station while keeping 
sufficient accuracy for noise reduction.

The laboratory and real condition tests provide data to optimize
the RFI cleaning for the next generation of  the AERA Front-End based on
Cyclone$\textsuperscript{\textregistered}$ V with Hardcore Processor System (HPS)
and System on Chip (SoC). 
\end{abstract}


\section{Introduction}
%
%
%
%
\IEEEPARstart{R}{ecently} 
radio detection of cosmic-ray air showers  relives a renaissance, mainly
thanks to a huge progress of the powerful digital signal
processing techniques in experiments such as LOPES \cite{LOPES}, 
CODALEMA \cite{CODALEMA} or the Auger Engineering Radio Array (AERA) 
\cite{AERA}, which is situated within the Pierre Auger Observatory (PAO) \cite{PAO}.

\begin{figure*}
\begin{centering}
\includegraphics[width=2.0\columnwidth,height=0.45\columnwidth]{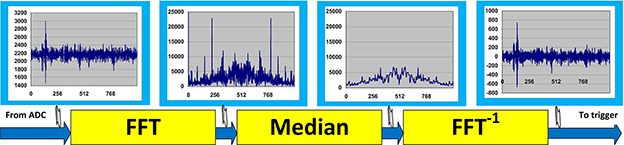}
\caption{A diagram showing a (FFT + Median filter + iFFT) chain cleaning the signal from the RFI contamination. 
The 1st graph shows the ADC input as unsigned data with an offset of ca. 2300 ADC-counts,
the 2nd - the absolute values of FFT coefficients in the frequency domain, the 3rd - FFT coefficients
“decontamined” by the median filter and 4th - signal converted back to the time domain. 
Additionally, the 0th FFT coefficient has been zeroed. Thus, the
cleaned signal in the time domain is represented as signed data without the offset. 
The amplitude of the signal remains roughly the same and the noise is considerably reduced.
}
\label{FFT_Median_iFFT}
\end{centering}
\end{figure*}

Results from the Pierre Auger Observatory,
point to the need for very large aperture detection systems
for ultra-high energy cosmic rays. With its nearly 100\% duty
cycle, its high angular resolution, and its sensitivity to the longitudinal
air-shower evolution, the radio technique is particularly well-suited 
for detection of ultra-high energy cosmic rays (UHECRs) in large-scale arrays.
AERA has been enlarged to 124 radio detector stations (RDSs), covering an
area of 6.5 km$^2$ therefore allowing the detection of UHECRs.

Since the 1960s we know that the radio emission from air showers is strongly correlated
with the local geomagnetic field \cite{Fegan}. In addition to the geomagnetic effect that 
can be described macroscopically \cite{Kahn}, Askaryan \cite{Askarian_A} \cite{Askarian_B} 
predicted that there should be an emission component related to the time-variation 
of the negative net charge excess in air showers. Both \cite{LOPES} and \cite{CODALEMA} confirmed these effects.
Using the AERA setup, the Auger collaboration quantified the relative strength of these effects \cite{AERA2}
The observation of air showers with radio-detection techniques can be done at almost all times. Moreover, radio signals
are sensitive to the development of the electromagnetic component of particle showers in the atmosphere
of the Earth. In the last 10 years the radio-detection technique in the MHz region has been revived and the
present radio-detector arrays for cosmic-ray research are equipped with low-noise and high-rate digital samplers.
Simultaneously, the number of stations within these arrays has grown from less than ten to more than one
thousand. The question to be addressed in the VHF band (MHz-range) is not whether extensive air showers emit
radiation. At this moment the main question to answer is: \emph{can we use radio signals to determine the primary
energy, the arrival direction, and the mass of cosmic rays with accuracies which are equal to or better than
those obtained by other techniques? And if yes, can we build for an affordable price a huge surface-detector
array based on the radio-detection technique?}

\section{RFI suppression for real AERA data}

Triggering directly on the radio signal of the air showers
(instead of using particle detectors as a trigger) poses
some challenges for the data acquisition, due to man-made
radio-frequency interference (RFI). The continuous background
level is set by the radio emission from the Galactic
plane, but any man-made narrow band transmitters add to
the level above which one must detect air-shower pulses.
Additionally, man-made pulsed RFI (from sparking electrical
equipment, airplanes, etc.) can mimic the short pulsed signal
from cosmic rays.
 Since the bandwidth and computational resources at each triggering level are
limited, one of the technical focuses for the first stage of
the array has been to develop various methods to reject
RFI in order to minimize efficiency losses from bandwidth
saturation.

The energy threshold of radio detection of cosmic rays is limited
by the considerable radio background and noise. The very
high level of RFI in the FM and short wave band has to be eliminated
by a band pass filter. Within the remaining receiver
bandwidth of 30 to 80 MHz the noise at the quiet-rural
environment of the Pierre Auger Observatory is dominated by
the frequency dependent galactic noise \cite{Galacticnoise} with noise temperatures
of 5000 K at 60 MHz

In addition to the galactic noise, there is a human made
background. This background consists of continuous signals,
as from a few radio and TV stations, and transients produced
by machines. Without an effective trigger, a stable and low level
energy threshold is not guaranteed. Furthermore, the data
rate for communication of the triggered data to the central
DAQ would exceed the available bandwidth.

\subsection{FFT + Median filter + iFFT}

For self-triggered measurements, the data will be digitized
and processed in real time by a powerful FPGA chip. The
narrow peaks in the frequency domain due to radio frequency
interferences have to be strongly suppressed before building
a trigger. These peaks are removed in several stations, digitizing at 180 MHz, 
using a median filter. The filter works in the frequency domain using the Fast Fourier
Transform (FFT) routine provided by Altera . Furthermore,
the phase of the signal deformed by the steep band pass filter
is reconstructed by a deconvolution in the frequency domain.
The median FPGA filter eliminates mono-frequent carriers,
but broadband radio pulses from cosmic showers are not
affected. After a second inverse FFT, signals are converted
back to the time domain. This chain of the digital signal
processing strongly enhances the signal to noise ratio, and
thus improves the radio pulse detection sensitivity (Fig. \ref{FFT_Median_iFFT}).

In order to suppress the strong man-made radio carrier signals below 
30\,MHz and above 80\,MHz and to fulfill the Nyquist theorem, the signals
of the antennas  are filtered using an analogue band filter
before being sampled by the ADCs.
The filters with the constant transfer function $H(f)$ are of high order with a non-constant group delay, 
thus leading to a dispersion and decreasing the amplitude of the input
signals $S(f)$. 
In a frequency domain, the resulting output signal $P(f)$ can simply be 
calculated using (\ref{conv}) :
\begin{equation}
P(f) = H(f) \cdot S(f)
\label{conv}
\end{equation}
Knowing $H(f)$ from measurements, it is now easy to get back $S(f)$ by 
inverting (\ref{conv}). This operation can be implemented into the
FPGA by placing a complex multiplication unit directly after the FFT
engine which multiplies the output data of the engine with the
precomputed coefficients of $1/H(f)$ stored in a RAM.
After transforming the signal back into the time domain,
the amplitude of the de-convoluted signal increases by about
20\% compared to the input signal. Since the galactic
and electronic noise is completely uncorrelated, the S/N ratio
increases by the same amount.

Aliasing appears when converted pulses are located close
to a border of converted blocks. It manifests by a spurious
contribution in the opposite border of the block and in the
neighboring block as well. This effect may cause spurious
triggers and has to be eliminated.

The problems can only be solved, without introducing dead
time between the blocks, by using an overlapping routine. 
Therefore the filter engine must run in another
clock domain with higher frequency (210 MHz instead of 180MHz).
Nevertheless, it requires additional resources e.g. FILO (First In Last Out)
memory based procedure to inverse in time a sequence of samples.
The FFT approach is generally very power consuming. This is a factor 
for a system supplied from solar panels. 

\subsection{IIR notch filter}

As stated earlier, before triggering on a radio pulse it is advantageous
to increase the signal-to-noise ratio by filtering out any
narrow band transmitters from the digitized antenna signals.
In stations digitizing at 200 MHz, this is accomplished in a computationally efficient
manner by using a series of infinite-impulse response (IIR)
notch filters in a Cyclone\textsuperscript{\textregistered}-IV FPGA. The IIR filters operate on the
time-domain signal, and the output of the filter $y_i$ is a linear
combination of input samples $x_j$ and delayed feedback
output samples $y_j$ from the filter:
\begin{eqnarray}
y_i = x_i & - & (2\cos \omega_N\cdot x_{i-1})  + x_{i-2} +   \nonumber \\
                 & + & (2r\cos \omega_N\cdot y_{i-1}) - (r^2\cdot y_{i-2})  
\label{eq:notch}
\end{eqnarray}
The normalized filter frequency $\omega_N = 2\pi f_N/f_S$ is given by the notch
frequency $f_N$ and the sampling frequency $f_S$.
The width parameter, r, is provided with a value strictly between 0 and
1, with higher values giving a narrower response function. For a narrow transmitter
a typical value is  r=0.99.
A complication in the implementation of the IIR filters in high-frequency FPGAs arises from the fact that one cannot arbitrarily
pipeline the feedback computation. We have resolved this by using the scattered look-ahead pipelining technique \cite{PARHI}, 
which increases the filter complexity but allows more time for computation in the FPGA.

\begin{figure}[b] 
\centering 
\includegraphics[width=8.1cm,keepaspectratio] {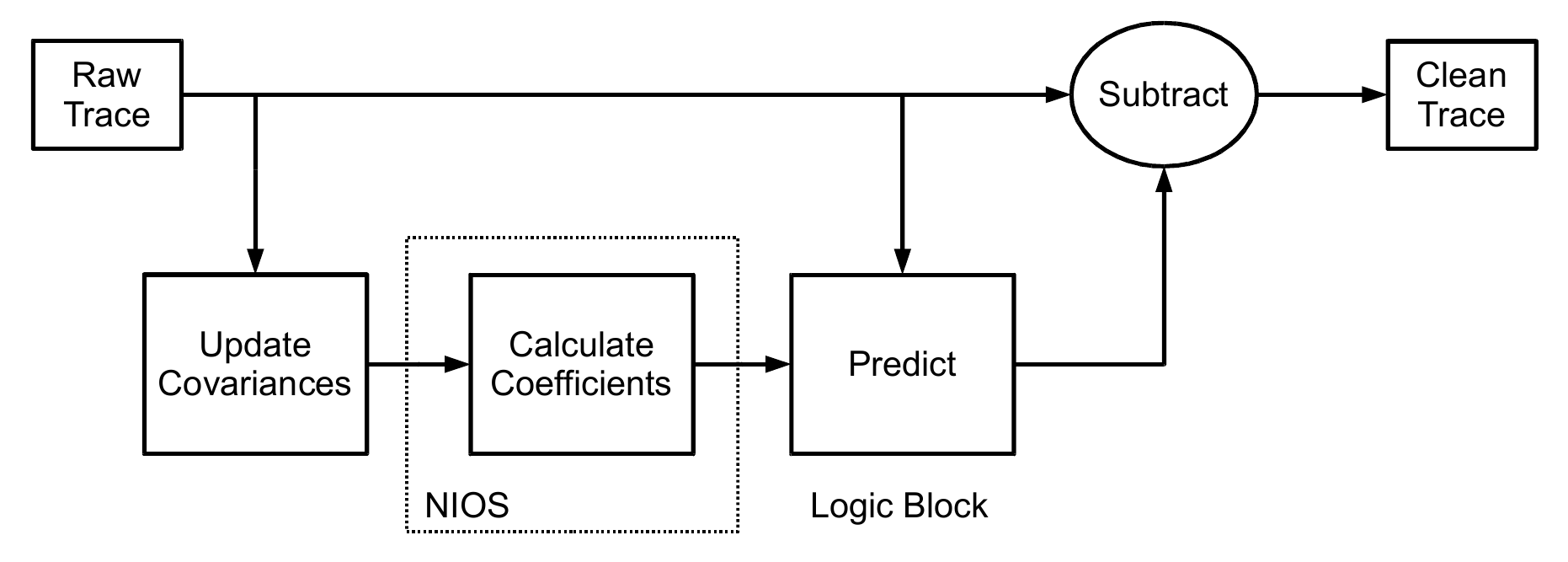}
\caption{The data flow of the FIR filter based on the LP method}
\label{LP_schematics}
\end{figure}

The coefficients of the $x_j$ and $y_j$ in Eq. \ref{eq:notch} can be precomputed for the desired notch frequencies, converted
to a fixed-point representation and loaded into the FPGA at run-time. The current design allows for four independent
tunable notch filters for each polarization direction and is applied on a Cyclone\textsuperscript{\textregistered} III, IV and V FPGA. 
It has been operating in the field for several years and successfully decreased and stabilized the threshold settings.

\begin{figure}[t]
\begin{centering}
\includegraphics[width=1\columnwidth]{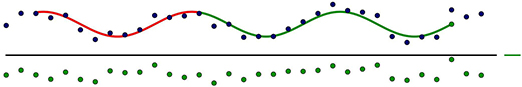}
\par\end{centering}
\vspace{5mm}
\begin{centering}
\includegraphics[width=0.895\columnwidth]{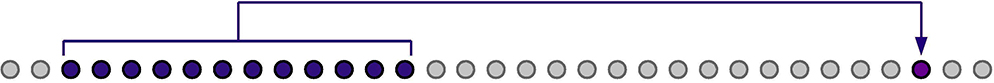}
{\footnotesize
\put(-209.15, -10){$\uparrow i-p-D$}
\put(-135.35,-10){$\uparrow i-D-1$}
\put(-21,-10){$\uparrow i$}}
\phantom{.....}
\par\end{centering}

\caption{\label{lp_D} An illustration of the method. 
The sine wave represents
the signal that is `fitted' (although in actuality no sinusoidal fit
is performed like this) where sample number $i$ is predicted by using
the samples $i-p-D$ to $i-D$. The predicted values are then subtracted
from the original values as illustrated by the green dots below the
horizontal bar with the minus sign, reducing the variance of the signal.}
\end{figure}

\subsection{Linear predictor}

Linear prediction is a mathematical operation where future values of a discrete-time signal 
are estimated as a linear function of previous samples \cite{Makhoul:LPReview}.
This method is widely used  in audio signal processing and speech processing for representing 
the spectral envelope of a digital signal of speech in compressed form, using the information 
of a linear predictive model \cite{Li}.
With the advent of faster signal processing techniques
in FPGAs it is now possible to apply similar 
techniques to the real-time processing of radio signals in the 
10 - 100 MHz region  \cite{IEEE_LP}. 

In the LP method  the covariances for 1024 ADC samples can be calculated in the 
FPGA fast logic block. Either the NIOS$\textsuperscript{\textregistered}$ processor, or the external ARM-processor, 
solves the matrix of 32 or 64 linear equations 
and provides coefficients needed for the FIR filter. The calculated 
coefficients are next transferred to the fast logic block, updating appropriate registers. 
They are used as the FIR coefficients in the ADC data filtering. Finally, 
the predicted and delayed data (expected background) are 
subtracted from the ADC data  to clean the signal from  
periodic contaminations (Fig. \ref{LP_schematics}).

Comparison of graphs in \cite{ICRC2013} indicates that the LP approach can eliminate
RFI narrow frequency contaminations as efficient as notch and FFT filters.

\begin{figure}
\begin{centering}
\includegraphics[width=1.0\columnwidth,height=0.52\columnwidth]{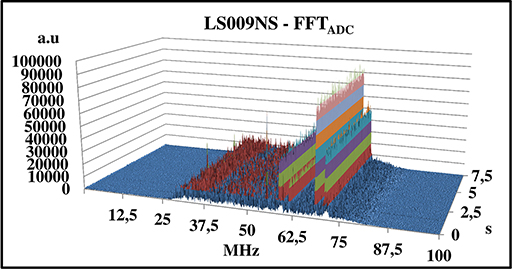}
\includegraphics[width=1.0\columnwidth,height=0.52\columnwidth]{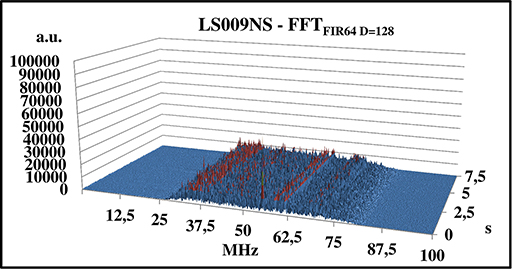}
\includegraphics[width=1.0\columnwidth,height=0.52\columnwidth]{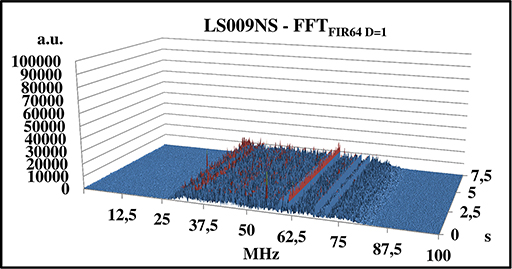}
\caption{FFT for ADC  and FIR filter with various D = 128 and 1, respectively, for long-term (7.5 s) data. 
Time 7.5 s corresponds to $1.5\cdot10^9$ cycles (sampling = 200 MSps).
Registered events contain 1024 samples.
The radio station LS009NS has been selected for a strong (100000 a.u. - arbitrary units) contamination by 4 carriers. 
The FIR64 filter shows a very good efficiency of the RFI suppression for all D factors (128 and 1).
}
\label{LS009NS}
\end{centering}
\end{figure}

\begin{figure}
\begin{centering}
\includegraphics[width=1.0\columnwidth,height=0.52\columnwidth]{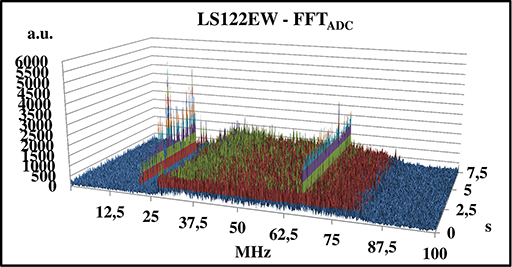}
\includegraphics[width=1.0\columnwidth,height=0.52\columnwidth]{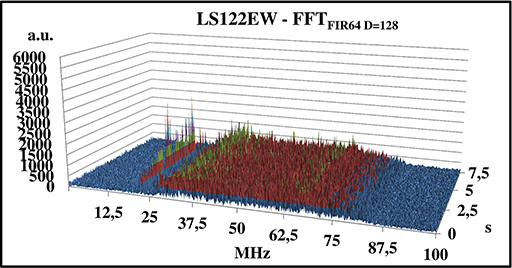}
\includegraphics[width=1.0\columnwidth,height=0.52\columnwidth]{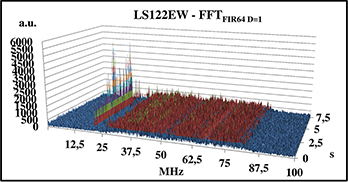}
\caption{Here, the radio station LS122EW has been selected for a weak (6000 a.u. - arbitrary units) contamination,
focused on two narrow bands around 12 and 62 MHz. The suppression for 62 MHz band is pretty good for
all D factors, however for the 12 MHz band, the RFI suppression is most efficient for D=128.
}
\label{LS122EW}
\end{centering}
\end{figure}

\begin{figure}
\begin{centering}
\includegraphics[width=1.0\columnwidth,height=0.53\columnwidth]{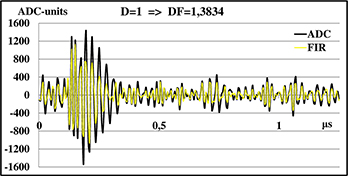}
\includegraphics[width=1.0\columnwidth,height=0.53\columnwidth]{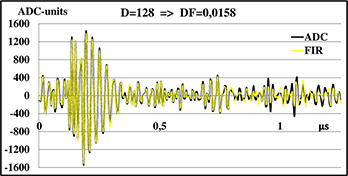}
\caption{Example of a potential signal distortion for D=1. It is recommended to introduce a delay
for a signal modification. For D=128 signals are almost unaffected.
 }
\label{various_D}
\end{centering}
\end{figure}

\begin{figure}
\begin{centering}
\includegraphics[width=1.0\columnwidth,height=0.55\columnwidth]{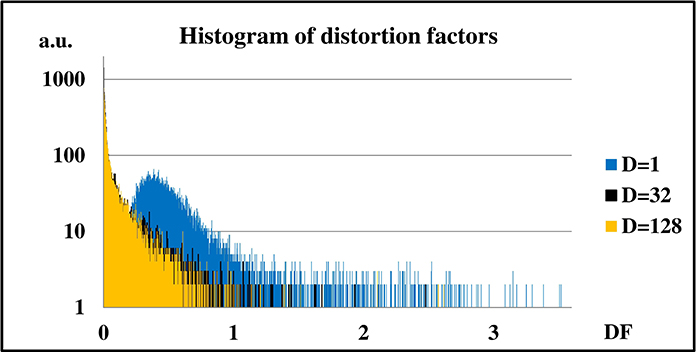}
\caption{Histogram of distortion factors (DF) for several hundreds AERA event filtered 
by the FIR filter with D=1, 32 and 128, respectively. It is visible that a configuration with D=1
distorts the signal too strong and cannot be used in a final design.
 }
\label{Fig_DF}
\end{centering}
\end{figure}

\section{Analysis of AERA events}

Figures \ref{LS009NS} - \ref{LS122EW} show spectra of original ADC traces and cleaned 
by the LP FIR filter with various values for parameter D of 128, 32 and 1, respectively (Fig. \ref{lp_D}). For the mono-carrier
contamination (Fig. \ref{LS009NS}) the RFI suppression is very good for any D parameter.
For more structured spectral contaminations the suppression depends on the D parameter ,
nevertheless, we observe a reduction of periodic noise.

The delay-line $D$ implies that there is a gap between the samples
that are used for the prediction and the sample that is to be predicted
(Fig. \ref{lp_D}). This delay-line is necessary to allow transient
signals to pass through the filter unaltered. For D=1 the efficiency of the 
RFI suppression is maximal, however,  the signal is significantly affected and 
a distortion factor (DF) (Fig. \ref{various_D}),
\begin{equation}
DF = \sum\limits_{k=-16}^{16}  \left(1 - \frac{(x_{FIR})_k}{(x_{ADC})_k} \right)^2
\label{DF}
\end{equation}
 introduced
to estimate a quality of filtering, reaches a large, unacceptable value. 
For laboratory tests we selected D=128, to keep a reasonable safety margin in measurements (Fig. \ref{Fig_DF}).

\begin{figure}[t]
\begin{centering}
\includegraphics[width=1.0\columnwidth,height=0.53\columnwidth]{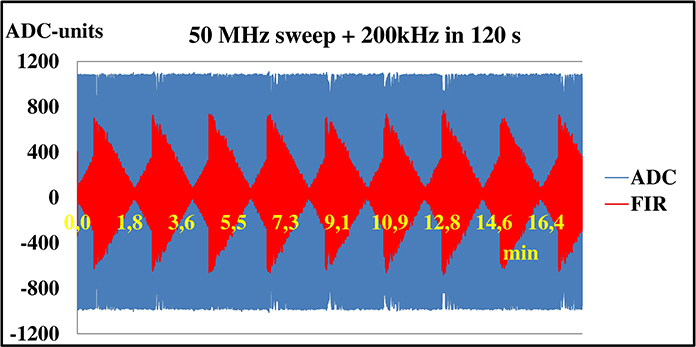}
\includegraphics[width=1.0\columnwidth,height=0.53\columnwidth]{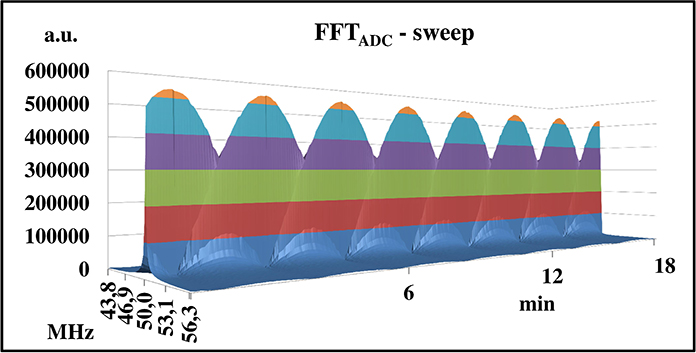}
\includegraphics[width=1.0\columnwidth,height=0.53\columnwidth]{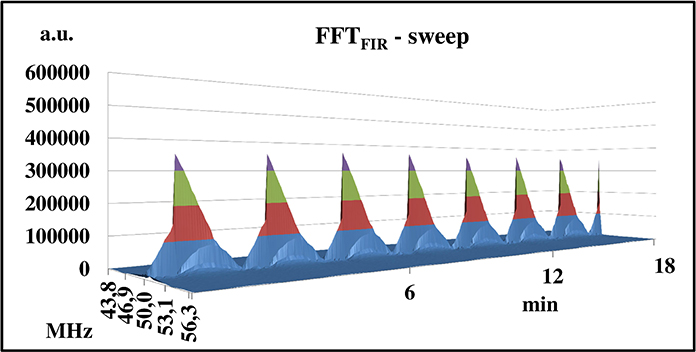}
\caption{Plots showing a suppression of the sine signal with a sweep from 50.0 MHz to 50.2 MHz in 2 minutes. 
For 8 such a cycles we observe a perfect repetition.
Plots obtained in real measurements on the laboratory setup  
for FIR64, D=128 and 14-bit LP coefficients.}
\label{sweep}
\end{centering}
\end{figure}

\section{Laboratory tests}

\begin{figure}
\begin{centering}
\includegraphics[width=1.0\columnwidth,height=0.65\columnwidth]{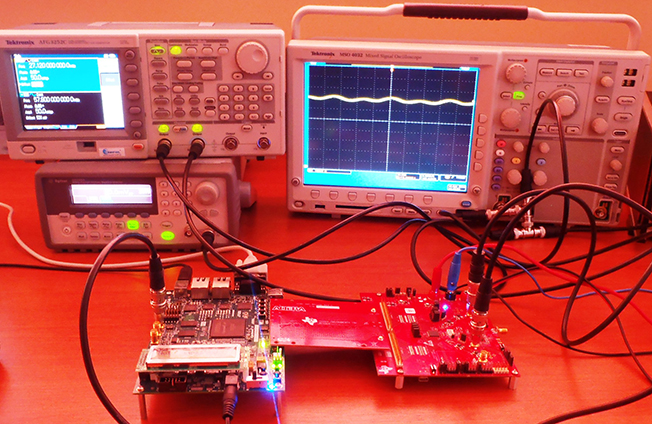}
\caption{The laboratory measurement setup with the Altera$\textsuperscript{\textregistered}$ 
DK-DEV-5CEA7N Development Kit with Cyclone$\textsuperscript{\textregistered}$ V FPGA and Texas Instr.
ADS4249EVM Evaluation Module with 2-channel 14-bits 250MSps ADC (ADS4249). Both modules were
connected by the Altera$\textsuperscript{\textregistered}$ HSMC-ADC-BRIDGE providing the LVDS data
transmission. Tektronix AFG3252C 2-channel generator provides 2 sine waveforms, Agilent 33250A 
provides noise signal. 
}
\label{lab_setup}
\end{centering}
\end{figure}

The LP FIR filter has been tested with the Altera$\textsuperscript{\textregistered}$ 
DK-DEV-5CEA7N Development Kit with a Cyclone$\textsuperscript{\textregistered}$ V FPGA and Texas Instr.
ADS4249EVM Evaluation Module with 2-channel 14-bits 250MSps ADC (ADS4249). Both modules were
connected through the Altera$\textsuperscript{\textregistered}$ HSMC-ADC-BRIDGE providing the LVDS data
transmission (Fig. \ref{lab_setup}).

\begin{figure}[t]
\begin{centering}
\includegraphics[width=1.0\columnwidth,height=0.53\columnwidth]{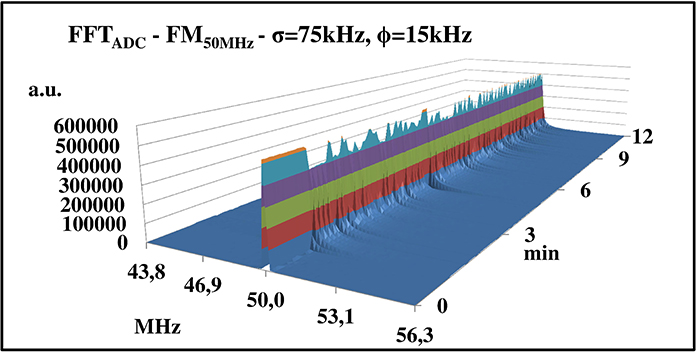}
\includegraphics[width=1.0\columnwidth,height=0.53\columnwidth]{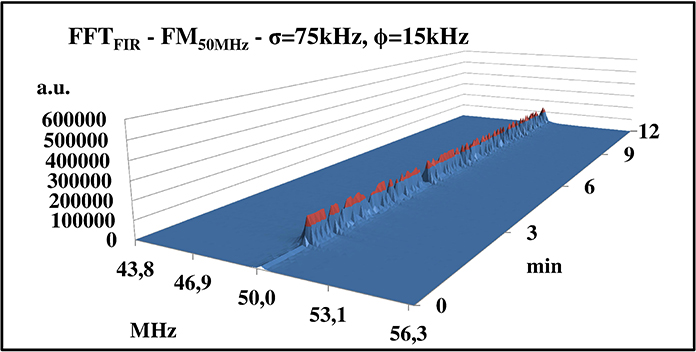}
\caption{Plots showing a suppression of the 50 MHz signal frequency modulated (FM) with a deviation of 
75 kHz and a modulation of 15 kHz. The signal corresponds to Hi-Fi radio transmission. Even for wide-band
FM the RFI suppression is relevant.  Plots obtained in real measurements on the laboratory setup  
for FIR64, D=128 and 14-bit LP coefficients.}
\label{FM}
\end{centering}
\end{figure}

At first the filter was tested by a mono-carrier drifting the frequency from 50.0 MHz to 50.2 MHz
in 120 s. The LP coefficients were not calculated by NIOS$\textsuperscript{\textregistered}$ in a correlation
with a generator running. Thus, these LP coefficients were used for the data cleaning in several tens of minutes
(which corresponds to more than $10^{12}$ clock cycles). Fig. \ref{sweep} shows a perfect long-term stability of the filter.
If the generator driving the filter uses the frequency for which the LP coefficients were originally calculated, the suppression is
almost total. Therefore, LP coefficients do not need to be refreshed very frequently.

\begin{figure}[t]
\begin{centering}
\includegraphics[width=1.0\columnwidth,height=0.53\columnwidth]{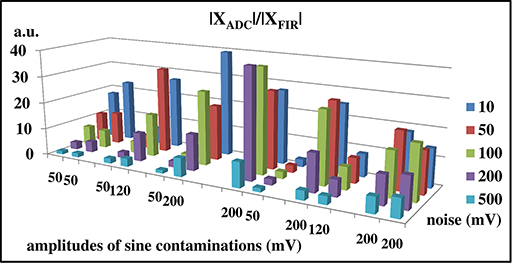}
\caption{Suppression factors for two sine pure carriers with frequencies 27.12 and 57.9 MHz and 
$V_{p-p}$ 50 and 200 mV, respectively, mixed with  10, 50, 100, 200 and 500 mV noise. }
\label{ratios}
\end{centering}
\end{figure}

\begin{figure}[t]
\begin{centering}
\includegraphics[width=1.0\columnwidth,height=0.53\columnwidth]{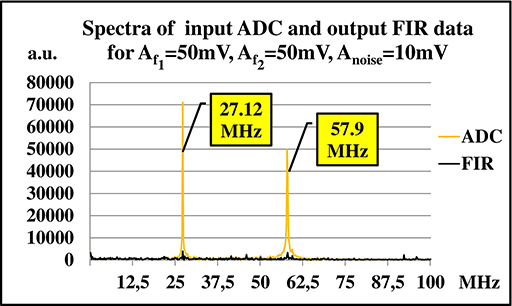}
\includegraphics[width=1.0\columnwidth,height=0.53\columnwidth]{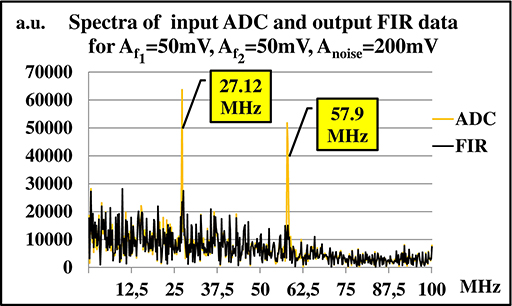}
\includegraphics[width=1.0\columnwidth,height=0.53\columnwidth]{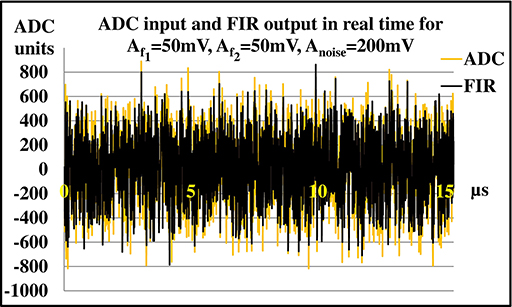}
\caption{Suppression factors for two sine pure carriers with frequencies 27.12 and 57.9 MHz and 
$V_{p-p}$ 50 and 200 mV, respectively, mixed with  10, 50, 100, 200 and 500 mV noise. Bottom graph
shows real time traces for 200 mV noise.  A suppression  in the time domain is almost invisible }
\label{ratios_with_noise}
\end{centering}
\end{figure}

Secondly, we checked the Hi-Fi FM configuration (75 kHz deviation of the 50 MHz carrier with the maximal acoustic
15 kHz modulation). This is very restrictive condition, which actually should not appear in real conditions in Argentina. 
The band 30 - 80 MHz is used rather by narrow-band transmitter, while the FM Hi-Fi transmission is selected 
for the band of 88 - 108 MHz, cut-off by the band-pass analog filter. Nevertheless, even these critical contaminations 
were successfully suppressed (Fig. \ref{FM}).

We also tested a suppression efficiency (in the Fourier space) in the presence of relatively strong white noise
(Fig. \ref{ratios_with_noise}). Two pure carriers with
27.12 and 57.9 MHz were wired mixed with noise. Fig. \ref{ratios} shows that even when the noise level reaches the signal level 
the suppression factor remains on a level of 5-10 (for signals 50 mV and 200 mV contaminated by 50 mV and 200 mV of a noise).
When the noise is small, the suppression factor reaches values up to 35. However, a strong asymmetry is observed. The strong
signal (i.e. 200 mV of 27.12 MHz) is suppressed with a very high factor (35) while a 4 times smaller signal (57.9 MHz)
is almost not suppressed at all. For vice versa configuration a suppression structure remain the same.

\section{Comparison to currently used IIR-notch filter}

\begin{figure}[h]
\begin{centering}
\includegraphics[width=1.0\columnwidth,height=0.55\columnwidth]{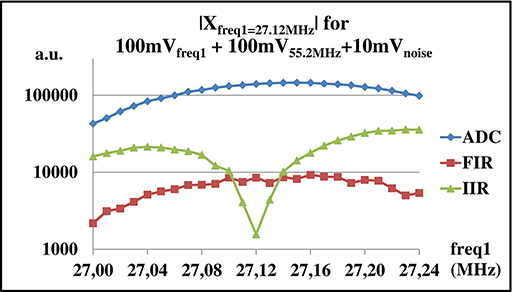}
\includegraphics[width=1.0\columnwidth,height=0.55\columnwidth]{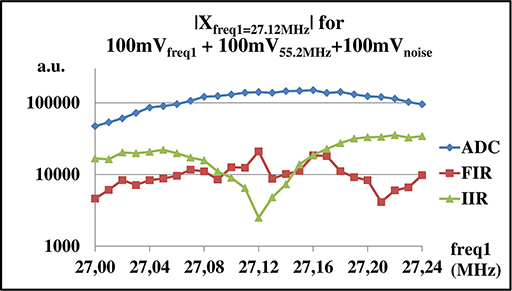}
\caption{Amplitudes of Fourier modules ($\bar{X}_{freq_1 = 27.12 MHz}$) for FIR filter based 
on the linear predictor and IIR-notch filter. The width of frequency bin is $\sim$0.1 MHz (1024-point FFT),
all $\bar{X}_{freq_1}$ contributions fall into $138^{th}$ frequency bin. The signal with various frequencies
$freq_1$ = 27.00 - 27.24 with a grid of 10 kHz is provided by the $1^{st}$ channel of Tektronic AFG3252C,
the $2^{nd}$ signal from the $2^{nd}$ channel of AFG3252C has a fixed frequency = 55.2 MHz, exactly equals
the reject frequency of he $3^{rd}$ stage of the IIR-notch filter. The reject frequency 
of the $1^{st}$ IIR-not filter is 27.12 MHz. Additional 10 mV or 100 mV noise is provided by Agilent 33250A generator. }
\label{FIR_IIR}
\end{centering}
\end{figure}

We compared suppression characteristics of the FIR filter based on the linear prediction with
the currently used IIR-notch filter with 4 band-reject bands. Fig. \ref{FIR_IIR} shows amplitudes of
the $138^{th}$ frequency bin in the Fourier space corresponding to a contribution of $1^{st}$ carrier contamination 
with a various frequencies in a range of 27.00 - 27.24 MHz for a small (10 mV - upper graph) and significant 
(100 mV - lower graph) noise. It is well visible that the efficiency of the IIR filter is very high and, as expected, 
only around the reject frequency (27.12 MHz) of the filter. Higher noise (comparison 100 mV vs. 10 mV)
reduces the efficiency only slightly, thereby extending the width of the rejection band (from $\sim\pm$20 kHz
to $\sim\pm$30 kHz). 

\begin{figure}[t]
\begin{centering}
\includegraphics[width=1.0\columnwidth,height=0.55\columnwidth]{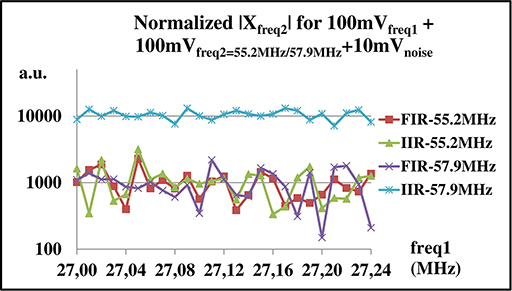}
\caption{Amplitudes of Fourier modules ($\bar{X}_{freq_2 = 55.2/57.9 MHz}$)  for the $2^{nd}$
mono-carrier contamination with 55.2 or 55.9 MHz, respectively. }
\label{freq1_freq2}
\end{centering}
\end{figure}

\begin{figure}[t]
\begin{centering}
\includegraphics[width=1.0\columnwidth,height=0.55\columnwidth]{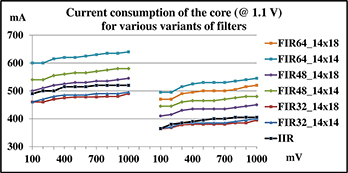}
\caption{The current consumption for several tested variants of the FIR filter with a comparison to the currently used IIR one. }
\label{power}
\end{centering}
\end{figure}

\begin{figure}[t]
\begin{centering}
\includegraphics[width=1.0\columnwidth,height=1.0\columnwidth]{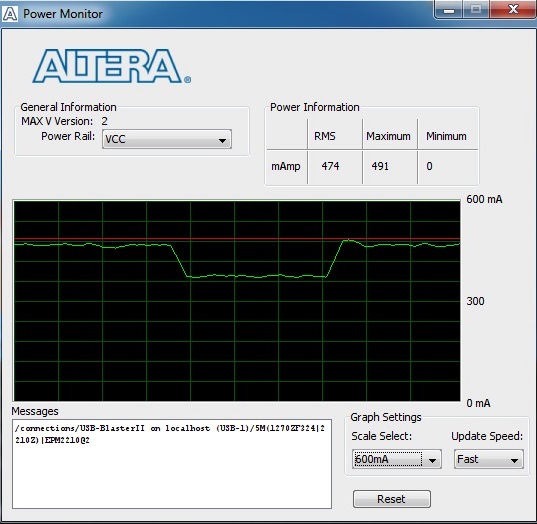}
\caption{A graph showing the current consumption of the core (@1.1 V). A reduction of the 
current consumption corresponds to temporary disabling of the NIOS processor }
\label{power_monitor}
\end{centering}
\end{figure}

\begin{figure}
\begin{centering}
\includegraphics[width=1.0\columnwidth,height=0.55\columnwidth]{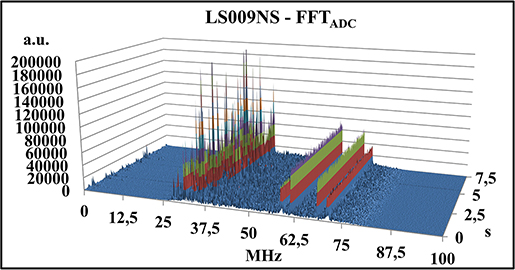}
\includegraphics[width=1.0\columnwidth,height=0.55\columnwidth]{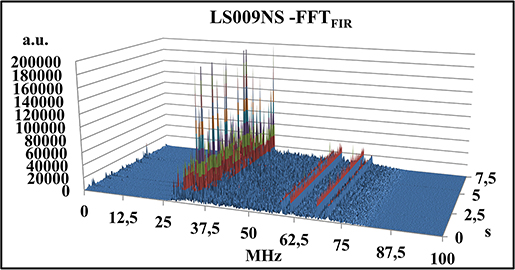}
\includegraphics[width=1.0\columnwidth,height=0.55\columnwidth]{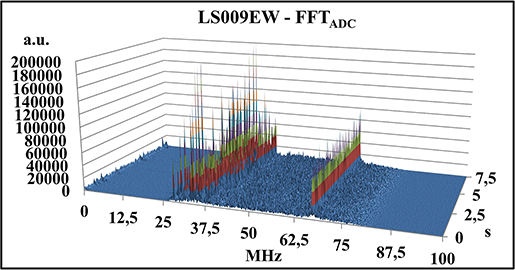}
\includegraphics[width=1.0\columnwidth,height=0.55\columnwidth]{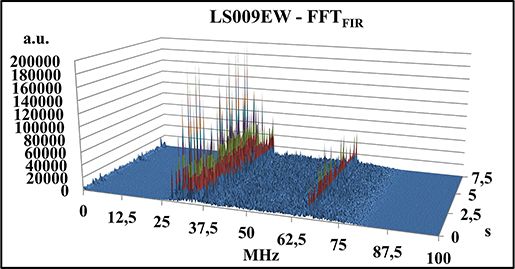}
\caption{ FFT for ADC  and FIR filter with D = 128, for long-term (7.5 s) data for the radio station LS009 and NS and EW polarizations. 
As for Fig. \ref{LS009NS} we see 4 mono-carriers, however with additional low frequency non-stationary RFI.
The presence of the variable in time RFI reduces an efficiency of the LP filter.
}
\label{pampas_LS009}
\end{centering}
\end{figure}

Fig. \ref{freq1_freq2} shows the $283^{th}$ (55.2 MHz) and $296^{th}$ (57.9 MHz) Fourier modules vs. 
various frequencies in the $1^{st}$ channel contamination (27.00 - 27.24 MHz). As expected, 
the suppression of the IIR filter outside the reject band is negligible. Note that for structured contaminations
(two mono-carriers with frequencies equal exactly the reject frequencies of the IIR-filter)
the efficiency of the FIR filter is comparable with the efficiency of the IIR one, provided the background 
noise level is low in comparison to the contamination. 

Figures \ref{FIR_IIR} and \ref{freq1_freq2} justify testing the FIR filter based on the linear predictor
as an adaptive filter, adjusting the suppression characteristics to changing RFI conditions in the field.

\section{Power consumption}

A power consumption is the factor for systems supplied from solar panels. More sophisticated 
filter provides a better accuracy of data processing. However, the power efficiency may significantly decreased.
We measured a power consumption for all developed FIR variants and compared with the currently
used IIR filter. Fig.\ref{power} shows results for the current consumption with the NIOS processor 
(left panel) and with temporary blocked NIOS (right panel). The NIOS processor was used for the
calculation of LP coefficients for the FIR filter and for data transmission via UART to the PC 
for both filters Fig.\ref{power_monitor}.

From Fig.\ref{power} it is visible that the FIR filter wit 32 stages and only 14-bit coefficients is at least
power efficient in comparable to the IIR one. Longer FIR filters consume much more power and could be used
exceptionally in extremely contaminated environment, where the suppression efficiency becomes the more important factor
than the power efficiency.

\section{Data from pampas}

We have implemented the 32-stage LP filter (with D=128) in the AERA radio station LS009.  
Fig. \ref{pampas_LS009} show a suppression of the RFI in the higher frequency range.
In comparison to Fig. \ref{LS009NS}b the suppression efficiency is lower, however, in current environment
the huge contribution of non-stationary RFI was observed. In a presence of variable RFI the efficiency of the LP
filter is not so high as for mono-carriers contamination.
Fig. \ref{pampas_LS009} shows that the noise suppression for the EW polarization is almost negligible.

\section{Conclusions}

Analytical calculations (Fig. \ref{LS009NS}-\ref{LS122EW}) show that a very high efficiency of the LP filter can be obtained.
The laboratory measurements confirmed that when the data is contaminated by mono-carriers, the suppression factor is very high.
However, the environmental RFI as encountered in the Pierre Auger Observatory has a much more sophisticated structure. Therefore, 
the efficiency of the LP-filter significantly decreases.

Nevertheless, in each radio station the filter setup can be calculated, thereby  
optimizing the suppression factor, depending on the location of each station and the type of
RFI contamination.

In AERA the Cyclone$\textsuperscript{\textregistered}$ IV FPGA EP4CE75F29I7 is currently in use.
The laboratory tests provide input for an optimization of
the RFI cleaning for the next generation of  the AERA Front-End based on
Cyclone$\textsuperscript{\textregistered}$ V with Hardcore Processor System (HPS)
and System on Chip (SoC).

\section*{Acknowledgment}
This work was supported by the Polish National Center for
Research and Development under NCBiR Grant No. ERA/NET/ASPERA/02/11,
by the National Science Centre (Poland) under NCN Grant No. 2013/08/M/ST9/00322,
, and by the Ministerie van Onderwijs, Cultuur
en Wetenschap, Nederlandse Organisatie voor Wetenschappelijk Onderzoek (NWO), Stichting voor Fundamenteel Onderzoek der Materie (FOM), 
the Netherlands.



\begin{thebibliography}{2}

\bibitem{LOPES} H. Falcke,W. D. Apel, A. F. Badea, et al.,
"Detection and imaging of atmospheric radio flashes from cosmic ray air showers",
\emph{Nature}, vol. {\bf435}, pp. 313-316, May 2005.

\bibitem{CODALEMA} D. Ardouin, A. Bell´etoile, D. Charrier, et al.,
"Radioelectric field features of extensive air showers observed with CODALEMA"
\emph{Astropart. Phys.}, vol. {\bf26}, pp. 341-350, Dec. 2006.

\bibitem{AERA} S. Fliescher for the Pierre Auger Collaboration, 
"Radio detection of cosmic ray induced air showers at the Pierre Auger Observatory", 
\emph{Nucl. Instr. Meth.}, ser. A, vol. {\bf662}, pp. 124-129, Jan. 2012.

\bibitem{PAO} J. Abraham et al., [Pierre Auger Collaboration],
``Properties and Performance of the Prototype Instrument for  the Pierre Auger Observatory'',
\emph{Nucl. Instr. Meth.}, ser. A, vol. 523, pp. 50-95, May 2004.

\bibitem{Fegan} D. J. Fegan, "Detection of elusive radio and optical emission from cosmic-ray showers in the 1960s",
\emph{Nucl. Instr. Meth.}, ser. A, vol. {\bf662}, pp. 2-11, Jan. 2012.

\bibitem{Kahn} F. D. Kahn and I. Lerche, "Radiation from Cosmic Ray Air Showers",
\emph{Proc. Roy. Soc. A}, vol. {\bf289}, pp. 206-213, Jan. 1966.

\bibitem{Askarian_A} G. A. Askaryan,
\emph{Journal of Exp. and Theoretical Phys.},
vol. {\bf14}, pp. 441, 1962.

\bibitem{Askarian_B} G. A. Askaryan, "Coherent Radio Emission from Cosmic Showers in Air and in Dense Media",
\emph{Journal of Exp. and Theoretical Phys.},
vol. {\bf21}, pp. 658-659, Jan. 1965.

\bibitem{AERA2} A. Aab et al., "Probing the radio emission from air showers with polarization measurements",
 Phys. Rev. {\bf D89}, 052002, Mar. 2014.

\bibitem{Galacticnoise}
G. A. Dulk, W. C. Erickson, R. Manning, and J.-L. Bougeret, 
"Calibration of low-frequency radio telescopes using the galactic background radiation",
\emph{A\&A}, vol. {\bf 365}, pp. 294-300, Jan. 2001

\bibitem{PARHI} K. K. Parhi and D. G. Messerschmitt, 
\emph{IEEE Trans. on Acoustics, Speech, and Signal Processing}, vol. {\bf37}, pp. 1099-1117, 1989.

\bibitem{Makhoul:LPReview} J. Makhoul, ``Linear prediction: A tutorial review'' 
\emph{Proc. of the IEEE}, vol. {\bf63}, no. 4, pp. 561-580, Apr. 1975.

\bibitem{Li} 
 D. Li,  D. O'Shaughnessy, "Speech processing: a dynamic and optimization-oriented approach",
 pp. 41–48. ISBN 0-8247-4040-8 (2003)

\bibitem{IEEE_LP}
Z. Szadkowski, E.D. Fraenkel, A. M. van den Berg,
"FPGA/NIOS Implementation of an Adaptive FIR Filter Using Linear Prediction
to Reduce Narrow-Band RFI for Radio Detection of Cosmic Rays",
\emph{IEEE Trans. on Nucl. Science}, vol. {\bf60}, pp. 3483-3490, Oct. 2013.

\bibitem{VCI}
Z. Szadkowski, E.D. Fraenkel, D. G{\l}as, R. Legumina,
"An optimization of the FPGA/NIOS adaptive FIR filter using
linear prediction to reduce narrow band RFI for the next generation
ground-based ultra-high energy cosmic-ray experiment",
\emph{Nucl. Instr. Meth.}, ser. A, vol. {\bf732}, pp. 535-539, June 2013.

\bibitem{ICRC2013} Z. Szadkowski, Ad M. van den Berg, E.D. Fraenkel, D. G{\l}as, J. Kelley, C. Timmermans, T. Wijnen,
"Analysis of the efficiency of the filters suppressing the RFI being developed for the extension of AERA",
\emph{ 33nd International Cosmic Ray Conference - July 2013 - Rio de Janeiro, Brazil}.


\end{thebibliography}
\end{document}